\def\be{\begin{equation}}
\def\te{\end{equation}}

\def\bea{\begin{eqnarray}}
\def\nn{\nonumber}
\def\tea{\end{eqnarray}}

\newcommand{\kh}{\mbox{$\chi_{0}$}}

\newcommand{\me}{\int d{\h}_1 d{\h}_2}
\renewcommand{\mp}{\mbox{$\tilde{m}^{+}$}}

\newcommand{\MP}{\mbox{${M^{+}}^2 $}}
\newcommand{\MN}{\mbox{${M^{-}}^2$}}

\def\h{\eta}

\def\l{\lambda}
\def\m{\mu}
\def\n{\nu}

\def\p{\pi}

\def\bo{{\raise.15ex\hbox{\large$\Box$}}}

\def\TH{{\raise.2ex\hbox{$\displaystyle \bigodot$}\mskip-4.7mu \llap H \;}}
\def\face{{\raise.2ex\hbox{$\displaystyle \bigodot$}\mskip-2.2mu \llap {$\ddot
        \smile$}}}

\def\Hat#1{\rlap{\kern.10em$\widehat{\phantom G}$}#1}
\def\HAt#1{\rlap{\kern.05em$\widehat{\phantom G}$}#1}

\def\cap#1{\rlap{\kern.1em$\widehat{\phantom{G\vrule height.8em}}$}#1{}}
\def\Cap#1{\rlap{\kern.05em$\widehat{\phantom{G\vrule height.8em}}$}#1{}}

\def\leftrightarrowfill{$\mathsurround=0pt \mathord\leftarrow \mkern-6mu
        \cleaders\hbox{$\mkern-2mu \mathord- \mkern-2mu$}\hfill
        \mkern-6mu \mathord\rightarrow$}
\def\overleftrightarrow#1{\vbox{\ialign{##\crcr
        \leftrightarrowfill\crcr\noalign{\kern-1pt\nointerlineskip}
        $\hfil\displaystyle{#1}\hfil$\crcr}}}

\def\frac#1#2{{\textstyle{#1\over\vphantom2\smash{\raise.20ex
        \hbox{$\scriptstyle{#2}$}}}}}
\def\ha{\frac12}
\def\sfrac#1#2{{\vphantom1\smash{\lower.5ex\hbox{\small$#1$}}\over
        \vphantom1\smash{\raise.4ex\hbox{\small$#2$}}}}
\def\bfrac#1#2{{\vphantom1\smash{\lower.5ex\hbox{$#1$}}\over
        \vphantom1\smash{\raise.3ex\hbox{$#2$}}}}
\def\afrac#1#2{{\vphantom1\smash{\lower.5ex\hbox{$#1$}}\over#2}}

\catcode`@=11
\def\underline#1{\relax\ifmmode\@@underline#1\else
        $\@@underline{\hbox{#1}}$\relax\fi}
\catcode`@=12

\def\nis{\nointerlineskip}
\def\Abar{\vbox{\nis\moveright.33em\vbox{
        \hrule width.35em height.04em}\nis\kern.05em\hbox{$A$}}{}}
\def\Dbar{\vbox{\nis\moveright.20em\vbox{
        \hrule width.50em height.04em}\nis\kern.05em\hbox{$D$}}{}}
\def\Gbar{\vbox{\nis\moveright.20em\vbox{
        \hrule width.50em height.04em}\nis\kern.05em\hbox{$G$}}{}}
\def\mbar{\vbox{\nis\moveright.15em\vbox{
        \hrule width.60em height.04em}\nis\kern.05em\hbox{$m$}}{}}
\def\Rbar{\vbox{\nis\moveright.20em\vbox{
        \hrule width.50em height.04em}\nis\kern.05em\hbox{$R$}}{}}
\def\Vbar{\vbox{\nis\moveright.05em\vbox{
        \hrule width.60em height.04em}\nis\kern.05em\hbox{$V$}}{}}
\def\Xbar{\vbox{\nis\moveright.20em\vbox{
        \hrule width.60em height.04em}\nis\kern.05em\hbox{$X$}}{}}
\def\thetabar{\vbox{\nis\moveright.15em\vbox{
        \hrule width.30em height.04em}\nis\kern.05em\hbox{$\theta$}}{}}
\def\Lambdabar{\vbox{\nis\moveright.25em\vbox{
        \hrule width.35em height.04em}\nis\kern.05em\hbox{${\mit\Lambda}$}}{}}
\def\Sigmabar{\vbox{\nis\moveright.25em\vbox{
        \hrule width.50em height.04em}\nis\kern.05em\hbox{${\mit\Sigma}$}}{}}
\def\phibar{\vbox{\nis\moveright.18em\vbox{
        \hrule width.40em height.04em}\nis\kern.05em\hbox{$\phi$}}{}}
\def\chibar{\vbox{\nis\moveright.12em\vbox{
        \hrule width.40em height.04em}\nis\kern.05em\hbox{$\chi$}}{}}
\def\psibar{\vbox{\nis\moveright.23em\vbox{
        \hrule width.40em height.04em}\nis\kern.05em\hbox{$\psi$}}{}}
\def\debar{\vbox{\nis\moveright.18em\vbox{
        \hrule width.35em height.04em}\nis\kern.05em\hbox{$\partial$}}{}}
\def\delbar{\vbox{\nis\moveright.10em\vbox{
        \hrule width.63em height.04em}\nis\kern.05em\hbox{$\nabla$}}{}}

\newskip\humongous \humongous=0pt plus 1000pt minus 1000pt

\newif\ifdtup

\def\ha{{1\over 2}}
\def\be{\begin{equation}}
\def\te{\end{equation}}
\def\bea{\begin{eqnarray}}
\def\tea{\end{eqnarray}}

\documentstyle[12pt]{article}

\makeatletter                                                           

\def\section{\@startsection {section}{1}{\z@}{-1.5ex plus -.5ex
minus -.2ex}{1ex plus .2ex}{\large\bf}}                                 

\def\@thmcountersep{}                                                   

\long\def\@makecaption#1#2{\vskip 10pt
\setbox\@tempboxa\hbox{#1. #2}   
   \ifdim \wd\@tempboxa >\hsize   
       #1. #2\par                 
     \else                        
       \hbox to\hsize{\hfil\box\@tempboxa\hfil}                         
   \fi}                                                                 

\def\ps@headings{                                                       
 \def\@oddhead{\footnotesize\rm\hfill\runninghead\hfill}
 \def\@evenhead{\@oddhead}                                              
 \def\@oddfoot{\rm\hfill\thepage\hfill}\def\@evenfoot{\@oddfoot} }

\makeatother                                                            

\begin{document}

\title{Minisuperspace as a Quantum Open System}{}{}

\def\runninghead{HU, PAZ AND SINHA: MINISUPERSPACE AS A QUANTUM OPEN SYSTEM}
\author {
{\em B. L. Hu} \thanks
{Department of Physics, University of Maryland,  College Park, MD 20742, USA}
\and
{\em Juan Pablo Paz} \thanks
{T6 Theoretical Astrophysics, ms 288, LANL, Los Alamos, NM 87545, USA }
\and
{\em Sukanya Sinha} \thanks
{Instituto de Ciencias Nucleares, UNAM,
Circuito Exterior, Apartado Postal 70-543,
CU Mexico, DF 04510, Mexico.}}

\date{} 

\pagestyle{headings}                                                   
\flushbottom                                                           

\maketitle
\vspace{-10pt} 

\begin{abstract}
We trace the development  of ideas on
dissipative processes in chaotic cosmology and on minisuperspace
quantum cosmology from the
time Misner proposed  them to  current research. We show \\
1) how the effect of quantum processes like particle creation in the
early universe
can address the issues of the isotropy and homogeneity of the observed
universe,\\
2) how viewing minisuperspace as a quantum open system can address the
issue of the validity of such approximations customarily adopted
in quantum cosmology, and \\
3) how invoking statistical processes like decoherence and correlation
when considered together can help to establish a theory
of quantum fields in curved spacetime
as the semiclassical limit of quantum gravity.
\end{abstract}

{\it Dedicated to Professor Misner on the occasion of his
sixtieth birthday, June 1992.}
To appear in the Proceedings of a Symposium on {\it
Directions in General Relativity},
College Park, May 1993, Volume 1, edited by B. L. Hu, M. P. Ryan and
C. V. Vishveshwara (Cambridge University Press 1993)~~~~umdpp 93-60

\section{Introduction}

In the five years between 1967 and 1972, Charlie Misner made an idelible
mark in relativistic cosmology in three aspects.

First he introduced the idea of chaotic cosmology. In contrast to the
reigning standard model
of Friedmann-Lemaitre-Robertson-Walker universe where isotropy and
homogeneity are `put in by hand' from the beginning,
chaotic cosmology assumes that the universe can have arbitrary irregularities
initially. This is perhaps a more general and philosophically pleasing
assumption. To reconcile an irregular early universe with the
observed large scale smoothness of the present  universe, one has to
introduce physical mechanisms to dissipate away the anisotropies and
inhomogeneities. This is why dissipative processes are
essential to the implementation of the chaotic cosmology program.
Misner (1968) was the first to try out this program
in a Bianchi type-I universe with the neutrino viscosity at work
in the lepton era. Though this specific
process was found to be too weak to damp away the shear, the idea remains a
very
attractive one. As we will see, it can indeed be accomplished
with a more powerful process, that of vacuum particle production at the
Planck time.  The  philosophy of chaotic cosmology is similar to that
behind the inflationary cosmology of Guth (1981), where the initial conditions
are rendered insignificant by the evolutionary process, which, in this case
is inflation.

Second, the  chaotic cosmology program ushered in serious
studies of the dynamics of Bianchi universes.
This was exemplified by Misner's
elegant work on the mixmaster universe (1969a). (That was when one of
us first entered the scene.) The ingeneous use of pictorial representation
of the curvature potentials made the complicated dynamics of these
models easy to follow and opened the way to a systematic analaysis of
this important class of cosmology. (See Ryan and Shepley 1975.)
This development complements and furthers the
ongoing program of Lifshitz, Khalatnikov and Belinsky (1963, 1970)
where, in seeking the most general cosmological solutions of
Einstein's equations using rigorous applied mathematics techniques
(see also Eardley, Liang and Sachs 1971),  they
found the inhomogeneous mixmaster universe (the  `generalized' Kasner
solution) as representing a generic behavior near the
cosmological singularity. This also paralleled
the work of Ellis and MacCallum (1969) who, following Taub (1951)
and Heckmann and Schucking (1962), gave a detailed analysis of the
group-theoretical structure of the Bianchi cosmologies. Misner's
chaotic cosmology program provided the physical rationale for these studies.
It also prepared the ground for minisuperspace quantum cosmology (1969b, 1972).

Third, the establishment of the quantum cosmology program, which
includes the application of the ADM quantization (1962) to cosmological
spacetimes, and the adaptation of the Wheeler- DeWitt superspace formulation
of quantum gravity (Wheeler 1968, De Witt 1967) to minisuperspace cosmology.
The physical motivation was to understand `the issue of the initial state'
(Wheeler 1964), and, in particular, the quantum
effects of gravity on the cosmological singularity. It was  Misner
and his associates who started the first wave of activity in quantum cosmology
(Ryan, 1972; see also Kuchar 1971, Berger 1974, 1975).
The second wave, as we know, came with the work of Hartle and Hawking (1983),
and Vilenkin (1983) who, while formulating the problem in Euclidean path
integral terms, opened the question on the boundary conditions of the universe.

These three directions in cosmology initiated by Misner and his collaborators
in the early seventies --chaotic cosmology, dissipative processes
and minisuperspace quantum cosmology-- have developed in major
proportions in the
past two decades. A lot of current research work in these areas still carry
the clear imprint of his influence. The present authors have had the good
fortune to be influenced by his way of thinking. We want to trace out some
major developments since Misner wrote these seminal papers  and describe
the current issues in these areas, with more emphasis on their interconnections
from early universe quantum processes to quantum cosmology.

\subsection{Particle Creation as a Quantum Dissipative Process}

Misner (1968) and  Matzner and Misner (1972) showed that neutrino viscosity
in the lepton era ($ \approx 1 sec$ from the Big Bang) is in general
not strong enough to dissipate away the shear in the Bianchi universes.
It was Zel'dovich (1970)
who first suggested that vacuum particle creation from the changing
gravitational
field in anisotropic spacetimes may act as a  powerful dissipative source.
This created a real possibility for the idea of chaotic
cosmology to work, now relying on quantum field processes effective near
the Planck time ($10^{-43}$ sec  from the Big Bang). One needs a new
theory --- the theory of quantum fields in curved spacetime.
Actually this theory was just beginning to take shape through the work
of Parker (1966, 1969) and Sexl and Urbantke (1968) at about the same time
when Misner was working on the isotropy of the universe problem (1968).
It was later realized that
the reason why the Bianchi I universe and not the Robertson-Walker universe
produces a copious amount of particles is because it breaks the conformal
invariance of the theory (Parker 1973).
Zel'dovich used a simple dimensional analysis to explain why
the produced particles can strongly influence the dynamics of spacetime
at the Planck time. The details are actually  more involved than this,
because one needs to remove the divergences in the
energy-momentum tensor of the matter field before it can be used
as the source of
the Einstein equation to solve for the metric functions. This is what
has been called the `backreaction' problem.
Zel'dovich and Starobinsky (1971) were the first to attempt such a
calculation for conformal scalar fields in a Bianchi I universe. (See also
Lukash and Starobinsky 1974).
That was before the basic issues of particle creation processes
were fully understood (e.g., ambiguities in the choice of the vacuum
in curved space, see, Fulling 1973)
and the regularization techniques were perfected. But they managed to show
the viability of such processes.
It took several years (1974-77) before
the common ground between the different ways of
regularization (e.g., adiabatic, point-splitting, dimensional, zeta-function)
was reached and  a firmer
foundation for this new theory of semiclassical gravity was established.
Hawking's discovery (1974) of black hole radiance, and the confirmed
`legality' of trace anomaly (Capper and Duff 1975, Deser, Duff and Isham 1976)
certainly invigorated this endeavour. The theoretical importance
of the subject and the infusion of talents from general relativity, field
theory, and particle physics have made this field an important component of
contemporary theoretical physics (Birrell and Davies 1982).
It is, as we shall see, also an intermediate step (the low energy or
adiabatic limit) towards a theory of quantum gravity and quantum cosmology.

Vigorous calculation of the backreaction of particle creation with regularized
energy momentum sources began with the work of Hu and Parker (1977,1978)
using adiabatic regularization methods, and Hartle (1977), Fischetti et al
(1979) and Hartle and Hu (1979, 1980), using effective action methods.
Their results on the
effect of particle creation on anisotropy dissipation confirmed the
qualitative estimate of Zel'dovich, and lent strong support for the
viability of Misner's chaotic cosmology program.

\subsection{Entropy of Quantum Fields and Spacetime?}

These efforts in quantum field theory in curved spacetime and
dissipative processes in the early universe contained the germs for
two interesting directions of later development. Both involve quantum
field theory and statistical mechanics applied to cosmological problems.
They are \\

\noindent i) Dissipation in quantum fields as a statistical thermodynamic
proces
   s:\\
Can one associate a viscosity with vacuum particle creation (Zel'dovich 1979,
Hu 1984) like that of neutrino in the lepton era, or leptoquark boson decays
in the GUT era?
Can one define an entropy of quantum fields? (Hu and Pavon 1986, Hu and
Kandrup 1987, Kandrup 1988)? An entropy of spacetime? (Penrose 1979, Hu 1983,
Davies, Ford and Page 1986)
Can one rephrase the chaotic cosmology idea in terms of some analog
of second law of thermodynamics involving both fields and geometry (Hu 1983)?
Can one generalize the fluctuation-dissipation relation (FDR)
for quantum processes
in spacetimes with event horizons (Candelas and Sciama 1977, Mottola 1986)
to general dynamical spacetimes? Indeed can one view the backreaction process
as
a manifestation of a generalized FDR  for dynamical systems like
quantum fields in cosmological spacetimes (Hu 1989, Hu and Sinha 1993)?

This first direction of research has been pursued in the eighties, with
the infusion of ideas and techniques in statistical and thermal field theory.
It is intended to be a generalization of thermodynamic ideas successfully
applied to
black hole mechanics (Davies 1978) to non-equilibrium quantum field systems
characteristic of cosmological problems. This was an offshoot of Misner's
interest in
dissipative processes in the early universe, only that one now takes on the
challenge of doing everything with quantum fields in non-equilibrium
conditions.
This exercise probes into many deeper issues of thermodynamics and cosmology,
such as the statistical nature of gravitational systems as different from
ordinary matter, the thermodynamic nature of spacetimes (with or without
event horizons), the nature of irreversibility and the cosmological origin
of time-asymmetry. We will not belabor this direction of research here, but
refer the reader to some recent reviews (Hu 1989, 1991, 1993a, 1993b).
As we will see, inquiries into the statistical nature of particle creation
and backreaction processes actually have a role to play in the second
direction, i.e.,

ii) Quantum field theory as the semiclassical limit of quantum
gravity. How good is modified Einstein's theory with backreaction from
quantum matter fields an approximation to quantum gravity?
Under what conditions can the quantum dynamics of spacetime described by a wave
function make the  transition to the classical picture described by
trajectories in superspace? We will use the paradigms of quantum open systems
applied to  Misner's minisuperspace models to illustrate these ideas.

\subsection{From Dissipation to Quantum Open Systems}

Speaking from the path traversed by one of us, one can identify
a few critical nodes in the search
for a pathway which connects i) particle
creation as a dissipative processes  to ii) backreaction  and semiclassical
gravity to iii) the statistical aspects in the issue of quantum to classical
transition
which underlies the relation of quantum gravity and semiclassical gravity.
(See the Ph.D. thesis of Sinha 1991 for an overview of other major
developments and references in these areas.)
The (in-out) effective action formalism (Schwinger 1951, DeWitt 1964, 1975)
was a correct start,
because it gives the rate of particle production and the backreaction
equations in a self-consistent manner (Hartle and Hu 1979). However the
conditions
for calculating the in-out vacuum persistence amplitude are unsuitable
for the consideration of backreactions, as the effective equations are
acausal and complex. The second major step was the recognition by
DeWitt (1986), Jordan (1986), and Calzetta and Hu (1987) that the in-in
or closed-time-path formalism of Schwinger (1961) and Keldysh (1964)
is the right
way to go. Calzetta and Hu (1987) derived the backreaction equations
for conformal scalar fields in Bianchi I universe, which provided
a beachhead for later discussion of the semiclassical limit of
quantum cosmology. They also recognized the capacity of this framework
in treating both the quantum and statistical aspects of dynamical fields
and indeed used it for a reformulation of non-equilibrium statistical
field theory (Calzetta and Hu 1988).
The third step in making the connection with statistical mechanics
is the construction of the so-called coarse-grained effective action
by Hu and Zhang (1990), similar in spirit
to the well-known projection operator
method in statistical mechanics (Zwanzig 1961).

This is where the connection with `quantum open systems' first
enters. The general idea behind the study of such systems is that one
starts with a closed system (universe) , splits it into a ``relevant" part
(system) and an ``irrelevant" (environment) part according to the
specific physical conditions of the problem (for example, the split
between light and heavy modes or slow and fast variables),
and follows the behavior only of the relevant variables (now
comprising an open system) by integrating out the irrelevant
variables. (See Hu 1993a and 1993b for a discussion of the
meaningfulness of such splits in the general physical
and cosmological context respectively).
Thus the coarse-grained effective action generalizes the background field/
fluctuation-field splitting  usually assumed when carrying out a quantum
loop expansion in the ordinary effective action calculation to that between
a system field and an environment field.
It takes into account the averaged effect of the environment variables
on the system and produces an effective equation of motion for the system
variables. This method was used by Hu and Zhang (1990) (see also Hu 1991)
to study the effect
of coarse-graining the higher modes in stochastic inflation, and by Sinha
and Hu (1991) to study the validity of minisuperspace approximations in
quantum cosmology. The last step in completing this tour
is the recognition by Hu, Paz and Zhang (1992)
that the
coarse-grained closed-time-path effective action formalism is equivalent
to the Feynman-Vernon (1963) influence functional which was
popularized
by Caldeira and Leggett (1983) in their study of quantum macroscopic
dissipative processes. Paz and Sinha (1991, 1992) have
used this paradigm for a detailed analysis of the relation of
quantum cosmology with semiclassical gravity.  We like to illustrate this
development with two problems, initially proposed and studied by Misner.
One is on the validity of minisuperspace approximation,
the other is on the classical limit of quantum cosmology.

\section{Validity of the Minisuperspace Approximation}
\setcounter{equation}{0}

In quantum cosmology, the principal object of interest is the wave function
of the universe $\Psi(h_{ij}, \phi)$, which is a functional on superspace. It
obeys the Wheeler DeWitt equation, which is an infinite-dimensional
partial differential equation. Its solution is a formidable task in the
general case. To make the problem more tractable, Misner suggested looking
at only a finite number of degrees of freedom (usually
obtained by imposing a symmetry requirement), and coined the word
minisuperspace (Misner 1972) quantization.
We note that such a tremendous simplification has its
preconditions. Since in the process of this transition one is
truncating an infinite number of modes, nonlinear interactions
of the modes among themselves and with the minisuperspace degrees of freedom
are being ignored. It is also well-known that this truncation violates
the uncertainty principle, since it implies setting the amplitudes and
momenta of the inhomogeneous modes simultaneously to zero. It is
therefore important
to understand under what conditions it
is reasonable to consider an autonomous evolution of the minisuperspace
wave function ignoring the truncated degrees of freedom. Misner was keenly
aware of the problems when he introduced this approximation.
The first attempt to actually
assess the validity of the minisuperspace approximation was made only
recently by Kuchar and Ryan (1986, 1989).

Two of us (Sinha and Hu 1991)
have tried to address this question in the context of interacting
field theory with the help of the coarse-grained effective action method
mentioned above. The model considered is that of a self interacting
(${\lambda \phi^{4}}$) scalar field coupled to a closed Robertson
Walker background spacetime. The ``system''(minisuperspace)
 consists of the scale factor
$a$ and the lowest mode of the scalar field, while the ``environment" consists
of the inhomogeneous modes. \footnote{In this example of minisuperspace the
scalar field should not be thought of as providing a matter
source for the Robertson-Walker background metric, since
in that case varying the action with respect to the scale factor will
not give the full set of Einstein equations. In particular, the $G_{ij}
= 8 \pi G T_{ij}$ equations that  constrain the energy momentum tensor
via $T_{ij} = 0$ for $i \neq j$ will be missing.}
In this model calculation the scalar field is mimicking
the inhomogeneous gravitational degrees of freedom in superspace.
We are motivated to do this
because linearized gravitational perturbations (gravitons)
in a special gauge can
be shown to be equivalent to a pair of minimally coupled
scalar fields (Lifshitz 1946).
Quantum cosmology of similar models of gravitational perturbations
and scalar fields have been studied by several authors (Halliwell and
Hawking 1985, Kiefer 1987, Vachaspati and Vilenkin 1988).

Our basic strategy will be to try to derive an ``effective"
Wheeler-DeWitt equation for the minisuperspace sector which contains
the ``averaged" effect of the higher modes as a backreaction term. We
will then explicitly calculate the backreaction term using the
effective action and consequently present a criterion for the validity
of the minisuperspace approximation.

\subsection{Effective Wheeler DeWitt Equation}
\setcounter{equation}{0}
The gravitational and matter actions in our model are given by
\bea
S_g & = & \ha \int d\eta a^{2}(1 - {a'^{2}\over a^2})\\
S_m & = & -\ha \int\sqrt{-g} d^{4}x[\Phi\Box\Phi +  m^2\Phi^2
          +{2\l\over 4!}\Phi^4 +  {R\over 6} \Phi^2]
\tea
where $a$ is the scale factor of a closed Robertson-Walker universe,
a factor
$l_p^2 = 2/(3\pi m_p^2) $ is included in the metric for simplification
of computations, and $\eta$ is the
conformal time.
$\Box$ is the Laplace Beltrami operator on the metric $g_{\m\n}$,
and $m$ is the mass of the  conformally coupled scalar field.
Defining a conformally related field $\chi = (al_p)\Phi$
and expanding $\chi$ in  scalar spherical harmonics $Q^{k}_{lm}(x)$
on the $S^3$ spatial sections,
\be
\chi = {\chi_{0}(t)\over (2\p^2)^{\ha}} + \sum_{klm}{ }f_{klm}
              Q^{k}_{lm}(x)
\te
where $k=2,3,\ldots \infty, l = 0,1,\ldots k-1, m = -l, -l+1 \ldots
l-1,l$ (henceforth we will use $k$ to denote the set $\{klm\}$).
We will make the further assumption that the interactions of orders
higher than quadratic of the
lowest(minisuperspace) mode $\chi_0$ with the higher modes $f_n$'s
($\sim \chi_0 f_{k}f_{l}f_{m}$) as well as the quartic self
interaction of the higher modes ($\sim f_{m}f_{n}f_{k}f_{l}$) are
small and can be neglected. With this assumption and with the
following redefinitions
$ m^{2} \rightarrow {m^{2}/l_p^2} , \l \rightarrow {\l/
2{\pi}^2}$,the matter action  can
be written as
\bea
S_m & = & \int d\eta \left\{{1\over 2} \left[ {\kh}'^2 \right.
 -m^2 a^2 {\kh}^2 \right] - {\l\over 4!}{\kh}^4
-{1\over 2}\sum_{k}{f_k \left[{d^2\over d{\eta}^2} + k^2 \right]
{f_k}}-{1\over 2}\sum_{k} {m^2 a^2 {f_k}^2}  \nn \\ \label{matter}
    &   &   -  {\l\over 4}\sum_{k}{{\chi_0}^2 {f_k}^2}
 \Big\}
\tea
The Hamiltonian constructed from the action $S = S_g + S_m$
is given by
\be
H = -{1\over 2}{\pi_a}^{2} + {1\over 2}{\pi_{\chi_0}}^{2} +
{1\over 2}\sum_{n}{\pi_{f_n}}^2
+ V_{0}(a, \chi_{0}) + V(a,\chi_{0}, f_n)
\te
where
\be
V_{0}(a,\chi_{0}) = -{1\over 2}a^2  + {1\over 2}m^{2}a^{2}{\chi_{0}}^2
+ {\l\over 4!}{\chi_0}^4
\te
and
\be
V(a,\chi_0 ,f_n)  =  {\l\over 4}\sum_{k}{\chi_0}^2 {f_k}^2
                    + {1\over 2}\sum_{k}(k^2 + m^2 a^2){f_k}^2
\te
where $\pi_a, \pi_{\chi_0},\pi_{f_n}$ are the momenta canonically
conjugate to $a, \chi_0, f_n$ respectively.
The Wheeler-DeWitt equation for the wave function of the Universe $\Psi$ is
obtained from the Hamiltonian constraint
by replacing the momenta by operators in  the
standard way, and is given by
\be
\left[{1\over 2}{\partial^2\over \partial a^{2}} - {1\over 2}
{\partial^2\over \partial{\chi_0}^{2}} - {1\over 2}\sum_{n}
{\partial^2\over \partial{f_n}^2} + V_{0} + V \right] \Psi(a, \chi_0, f_n) = 0
\te
where we choose a
factor ordering such that
the kinetic term appears as the Laplace-Beltrami operator on superspace.

Writing
\be
\Psi(a,\chi_0,f_n) = \Psi_0 (a, \chi_0) \Pi_n \Psi_n(a, \chi_0, f_n)
\te
we would like to obtain from (2.8) an effective Wheeler-DeWitt equation
of the form
\be
( H_0 + \Delta H ) \Psi_0 (a, \chi_0) = 0
\te
where $H_0$ is the part of the Hamiltonian operator in (2.8)
independent of $f_n$
and  $\Delta H$ represents the influence of the higher modes. By making
the assumption that  $\Psi_n$ varies slowly with the minisuperspace variables
(see Sinha and Hu 1991 for details of this approximation) one can identify
\be
\Delta H = -\sum_{n} <H_n>
\te
where the expectation value is taken with respect to $\Psi_n$.
It is evident that the examination of this term will  enable us to
comment on the validity of the minisuperspace description.
We will then make a further assumption that $\Psi_0$ has a WKB form,
i.e, $\Psi_0 = e^{iS(a,\chi_0)}$ which can be used in regions of
superspace where  the wavefunction oscillates rapidly, such that
using eqn. (2.10), $S$ satisfies
\be
{1\over 2} {(\nabla S)}^2 + V_0 = -\sum_{n}< H_n> \label{HJ}
\te
where $\nabla$ is the gradient operator on minisuperspace.
The above equation can be regarded as a Hamilton-Jacobi equation with
backreaction. It can be
also shown  that the $\Psi_n$'s satisfy a  Schr\"odinger - like
equation with respect to the WKB time.
This approximation can therefore be roughly thought of as the
semiclassical limit where the minisuperspace modes behave
classically, but the higher modes behave quantum mechanically ( for
further subtleties regarding the semiclassical limit see Sec. 3).
 Identifying ${\partial S\over \partial a} = \p_a$
and ${\partial S\over \partial\chi_0} = \pi_{\chi_0}$, and
substituting for the canonical momenta in  terms of ``velocities"
$a'$ and $\chi_0'$ Eq. (12) reduces to
\be
{1\over 2}a'^{2} -{1\over 2}{\chi'_0}^2 + V_0(a,\chi_0) = -\sum_n<H_n>
\te
This is the effective Wheeler- De Witt equation in the WKB limit,
which we will compare with the backreaction equation derived in the
next section, in order to calculate the term on the right hand side.

\subsection{Backreaction of the Inhomogeneous Modes}

\par
We would now like to calculate the backreaction term (2.11)
explicitly. Since we are making a split of the system from the environment
based
 on
the mode decomposition, we should use the coarse-grained effective action
where the coarse graining consists of functionally integrating out
the higher modes.
As we would like to generate vacuum expectation values from the
effective action rather than the  matrix elements one should
use the in-in or closed-time-path (CTP) version of the effective action
(Calzetta and Hu 1987) rather than the in-out version.
The CTP coarse-grained effective action in our case is given by
(for details, see Sinha and Hu 1991)
\be
e^{iS_{eff}( a^+ ,{\chi}_{0}^{+},a^-,{\chi}_{0}^{-}  )}  =
 \int {\cal D}f_{k}^{+}
{\cal D}f_{k}^{-} e^{i\left( S(a^+ ,{\chi_0}^+, f_{k}^{+})
- S^{*}( a^- ,{\chi_0}^{-} ,f_{k}^{-}) \right)}
\te
$S^{*}$ indicates that in this functional integral, $m^{2}$ carries
an $i\epsilon$  term.
$a^{\pm}, \chi_{0}^{\pm}, f_{k}^{\pm}$ are the fields in the
positive (negative) time branch running from $\eta =
\stackrel{-}{+}\infty$
to $\pm\infty$.
The path integral
is over field configurations that coincide at $t = \infty$.
${\cal D} {f_k}$ symbolizes the functional integration measure over the
amplitudes of the higher modes of the scalar field.

We have  derived the
one loop renormalized coarse-grained effective action given
as follows
(omitting terms that involve $-$ fields only)
\bea
S_{eff} & = & S_{g+} + {1\over 2}\int d\h\left\{ {{\kh^+}'}^2
- \mp^2 {\kh^+}^2\right\} \nn \\
        &   &-{\l\over 4!}\int d\h {\kh^+}^4 +  {13\l\over 48}\int d\h\MP
+ {1\over 16}\int d\h {M^+}^4 \nn \\
        &   &+ {1\over 32} \me \MP(\h_1)
K(\h_1 - \h_2) \MP(\h_2)\nn\\
&  & +{1\over 32}\me \MP(\h_1)
\bar{K}(\h_1 - \h_2)\MN(\h_2)
\tea
where the coupling constants have their renormalized values.
$S_{g_+}$ represents the classical gravitational part of the action.
$M^{{\pm}^2} = {\tilde{m}}^{{\pm}^2} + \ha \l \kh^{{\pm}^2} $, and
${ K}$ and ${\bar{K}}$ are complex nonlocal kernels with explicitly known
forms.
The  effective equations of motion are obtained
from this effective action via
\footnote{we need
to compute only those terms in the effective action that involve the +
fields. Terms containing only $-$ fields will not contribute to the
equations of motion.}

\be
{{\delta S_{eff}\over \delta a^{+}}\Big|}_{\stackrel{a^+ = a^- = a}{\chi_{0}
^{+} = \chi_{0}^{-} = \chi_{0}}} = 0 \quad {\rm and}\quad
{{\delta S_{eff}\over \delta\chi_{0}^{+}}\Big| }_{\stackrel{ a^+ = a^- = a}
{\chi_{0}^{+} = \chi_{0}^{-} = \chi_{0}}} = 0
\te

Since we are interested in comparing  with  (2.13) which is
equivalent to the $G_{00}$ Einstein equation with backreaction, we
need the first integral form of (2.16), which can be derived as
\bea
& & {1\over 2}{a'}^2 - {1\over 2}\kh'^2 + {1\over 2} m^2 a^2 \kh^2 - {1\over 2}
a^2  + {\l\over 4!}\kh^4
 -{13\l\over 96}\kh^2 - {13\over 48}m^2 a^2 - {1\over 16}
M^2 ln \m a \nn \\
& & + {1\over 32}\int d\h_1
M^2(\h){\cal K}(\h - \h_1)M^2(\h_1)
= 0
\tea
with the assumption of having no quanta of the higher modes in the
initial state.
${\cal K} = K + \bar{K}$ and is real and hence the above equation
is also manifestly real.
Equation (2.17) is then equivalent to the effective $G_{00}$
Einstein equation or the Einstein- Hamilton- Jacobi equation plus
backreaction. This in turn can be identified with  (2.13), the
effective  Wheeler- De Witt equation in the WKB limit. Therefore,
comparing
(2.17) and (2.13) we can identify the backreaction piece in (2.13) as
\bea
\sum_{n}<H_n> & = & -{13\l\over 96}\kh^2 - {13\l\over 48}m^2 a^2
-{1\over 16}M^2 ln \m a \nn \\
&  & - {1\over 32}\int d\h_1 M^2(\h)
{\cal K}(\h - \h_1)M^2(\h_1)
\tea
when the boundary conditions on the wave function are appropriate for
the $\Psi_n$'s to be in a conformal ``in" vacuum state. \footnote
{Since $\cal K$ is real the backreaction term given above is real and
 represents a genuine expectation value in the
``in" vacuum state rather than  an in-out matrix element generated using
the in-out effective action.}

Since equation (2.17) is  the ``effective"
Wheeler- DeWitt equation for the minisuperspace sector within our
approximation scheme, the condition for validity of this approximation
can be stated as
\be
\sum_{n}<H_n> \quad \ll \quad  V_0
\te
where by the left hand side we mean the regularized value given by
(2.18).
It was shown that the term in equation (2.18) involving the nonlocal kernel
is related to dissipative behavior in closely related models
(Calzetta and Hu 1989).
This dissipative behavior in turn has been related to particle production
by the dynamical background geometry in semiclassical gravity models
(Hu, 1989).
In our case this can be interpreted as scalar particles in the higher
modes being produced as a result of the dynamical evolution of the
minisuperspace
degrees of freedom generating a backreaction that modifies the
minisuperspace evolution.
 We can therefore think of this term as introducing dissipation
in the minisuperspace sector due to interaction with the higher modes
that are integrated out. One can justifiably think of autonomous
minisuperspace evolution only when this dissipation is small. Since
we have used the scalar field modes to simulate higher gravitational
modes these considerations can also be directly extended to include
gravitons.
\footnote{A similar idea has been discussed by Padmanabhan and Singh
(1990) in a linearized gravity model where they claim that in
order that the minisuperspace
approximation be valid, the rate of production of gravitons should be
small.}

This is an example of how ideas of open systems can be useful
in understanding dissipation and backreaction, even in quantum cosmology.
One can also use this paradigm to address the problem of quantum to classical
transition, specifically the relation of semiclassical gravity with
quantum gravity, which we will now address. To do this
the formalism will need to be elevated to the level of density matrices.

\section{Semiclassical Limit of Quantum Cosmology}
\setcounter{equation}{0}

We now report on the result of some recent work by two of us
(Paz and Sinha 1991, 1992) on this  problem.
Quantum cosmology rests on the rather bold hypothesis that the entire
universe can be described quantum mechanically. This pushes us to
question the usual Copenhagen interpretational scheme that relies
on the existence of an {\it a priori} classical external
observer/apparatus. Since in the case of quantum cosmology this
cannot be assumed, the theory needs to predict the
``emergence" of a quasiclassical domain starting from a fundamentally
quantum mechanical description. Recently, there has been a lot
of interest in this subject, and two basic criteria for classicality
have emerged (see Halliwell 1991 and references therein) from these
endeavors. The first is
{\it decoherence} -- which requires that quantum interference between
distinct alternatives must be suppressed. The second is that of
{\it correlation}, which requires that the wave function or some
distribution constructed from it (e.g. Wigner 1932) predicts
correlations between
coordinates and momenta in accordance with classical equations of
motion.  We will study the emergence of a semiclassical limit with the help
of a quantum open system paradigm applied in the context of quantum cosmology.
The basic mechanism for achieving decoherence and the
appearance of correlations is by coarse-graining certain variables
acting as an environment coupled to the system of interest.

\subsection{Reduced Density Matrix}
We will consider a $D$-dimensional minisuperspace with coordinates $r^m$
($m=1,\ldots,D$) as our system
 for which the
Hamiltonian can be written as:
\begin{equation}
H_g= {1\over{2M}}G^{mm'}p_mp_{m'} + M V(r^m)
\end{equation}
where $G^{mm'}$ is the (super)metric of the minisuperspace
and $M $ is related to the square of the Planck mass.
The minisuperspace modes are coupled to
other degrees of freedom $\Phi$, such as the modes of a scalar field,
or the gravitational wave modes with a
Hamiltonian $H_\Phi(\Phi, \pi_\Phi, r^m, p_m)$.
These constitute
the environment, or the `irrelevant' part in our model.
Thus, the wave function
of the Universe is a function $\Psi=\Psi(r^m, \Phi)$ which
satisfies the Wheeler-De Witt equation (Wheeler 1968,
DeWitt 1967) :
\begin{equation}
H\Psi= \bigl(H_g + H_\Phi\bigr)\Psi=0
\end{equation}
where the momenta are now replaced by operators and $H_{\Phi}$.
We use the same factor ordering as before.
If one is interested in making predictions only about the behavior of the
minisuperspace variables, a suitable quantity for such a
coarse-grained description is the reduced density matrix defined as:
\begin{equation}
\rho_{red}(r_2,r_1) = \int d\Phi \Psi^{*}(r_1,\Phi) \Psi(r_2,\Phi)
\end{equation}
We will consider an ansatz for the wave function of the following
form:
\begin{equation}
\Psi(r,\Phi) = \sum\limits_n e^{iMS_{(n)}(r)}C_{(n)}(r)\chi_{(n)}(r,\Phi).
\end{equation}
where $S_n$ is a solution of a Hamilton-Jacobi equation
with respect to the $r$ coordinate  which is the same as (2.12)
without the backreaction term and with the gradient given by
$G^{mm'}{{\partial }\over{\partial r^{m'}} }$. The prefactor $C_{(n)}
$is determined through the equation
$2G^{mm'} {\partial_mC_{(n)}} \partial_{m'}S_{(n)} + \Box_G S_0 = 0$,
and $\chi_{(n)}$ is a solution of a Schr\"odinger equation with
respect to the WKB time
${d\over dt} = G^{mm'}{{\partial S_0}\over{\partial r^{m'}} }
{{\partial}\over{\partial r^m}}$ and
with Hamiltonian $H_{\Phi}$. These equations are derived by the well
-known order by order expansion
in $M^{-1}$, where $M$ acts as a parameter analogous to the mass of
the heavy modes (in our case the minisuperspace ones) in the
Born-Oppenheimer approximation. The Hamilton-Jacobi equation
satisfied by $S_{(n)}$ will have a $D-1$ parameter family of
solutions, the subindex $(n)$ labeling such parameters
characterizes a particular solution and thus a specific WKB branch.

The reduced density matrix associated with the wave function $(3.4)$ is then
given by:
\begin{equation}
\rho_{red}(r_2,r_1) =\sum\limits_{n,n'} e^{iM[S_{(n)}(r_1) - S_{(n')}(r_2)]}
    C_{(n)}(r_1)C_{(n')}(r_2) I_{n,n'}(r_2,r_1)
\end{equation}
\noindent where we  call
\begin{equation}
I_{n,n'}(r_2,r_1)= \int \chi^{*}_{(n')}(r_2,\Phi)\chi_{(n)}(r_1,\Phi) d\Phi
\end{equation}
the influence functional.
The justification for this name comes from the fact that
it has been shown (Paz and Sinha 1991, Kiefer 1991)
that it is exactly analogous to
the Feynman-Vernon (1963) influence functional
in the case in which the environment is initially in a pure state.
For models with $r >1$ it can be shown that it is indeed a
functional of two histories, which is in turn related to the
Gell-Mann Hartle decoherence functional (Gell-Mann and Hartle 1990, Griffith
1984, Omnes 1988). It is the central object of our consideration.

\subsection{Decoherence and Correlation}
We will now study the emergence of classical behavior
in the minisuperspace variables as a consequence of their
interaction with the environment. As we will show the two basic
characteristics of classicality,
the issues of correlations and decoherence, are indeed interrelated.

{\bf i) Decoherence}:
Decoherence
occurs when there is no interference effect between alternative histories.
If each  of the diagonal terms in the sum of $(3.4)$
corresponds to a nearly classical set of histories, the
interference between them is contained in the non--diagonal terms ($n\neq n'$)
of that sum. Thus, the system decoheres if
 influence functional  $I_{n,n'}\propto\delta_{n,n'}$ approximately.
\footnote{for
a suggestive argument to justify the choice of $I_{n,n'}(r,r)$ as an indicator
of the degree of decoherence between the WKB branches, see Paz and Sinha (1992)
Sec. V}
As a consequence the density matrix $(3.4)$ can be considered to
describe a ``mixture'' of non--interfering WKB branches representing
distinct Universes.

{\bf ii) Correlations}:
The second criterion for classical behavior is the existence of
correlations between the coordinates and momenta which approximately
obey the classical equations of motion. To analyze this aspect we
will compute the Wigner function associated with the reduced density
matrix and examine whether the Wigner function has a peak about a
definite set of trajectories in phase space corresponding to the
above correlations (Halliwell 1987, Padmandabhan and Singh 1990).
Once we have established the decoherence between the WKB branches
using criterion i), we look for correlations using the Wigner
function {\it within} a ``decohered WKB branch", i.e, that associated with
$(n = n')$. In this sense the two criteria are interrelated, i.e, we
{\it need} decoherence between the distinct WKB branches to be able to
meaningfully predict correlations.

The Wigner function associated with one of the diagonal terms is
given by
\begin{equation}
W_{(n)}(r,P) = \int\limits_{-\infty}^{+\infty} d\xi^m
C_{(n)}(r_1)C_{(n)}(r_2)
e^{-2i{P_m\xi^m}}
e^{iM[S_{(n)}(r_1) - S_{(n)}(r_2)]}I_{n,n}(r_2,r_1)
\end{equation}
\noindent where  $r^m_{1,2} = r^m \pm {\xi^m\over M}$.

The functional $I_{n,n}(r_2,r_1)$
plays the dual role of producing the diagonalization of $\rho_{(n)}$
and affecting the correlations (the phase affects the correlations
and the absolute value the diagonalization).
\footnote{The diagonalization produced by $I_{n,n}(r_2,r_1)$ has been
studied
by various authors (Kiefer 1987, Halliwell 1989, Padmandabhan 1989,
Laflamme and Louko 1991) and
has been identified with decoherence. However, as noted by Paz and
Sinha (1992), this term applies more properly
to the lack of interference
between WKB branches which, as emphasized by Gell-Mann and Hartle (1990),
is the more relevant effect. However, it is clear that
the diagonalization is an effect that accompanies the former one and
has the same origin.}
In the language of measurement theory (Wheeler and Zurek 1986)
one can say that the environment is ``continuously
measuring'' the minisuperspace variables  (Zurek 1981, 1991, Zeh 1986,
Kiefer 1987) and that this interaction not only suppresses the
$n\neq n'$ terms in the sum of equation $(3.11)$, but also
generates a ``localization'' of the $r$
variables inside each WKB branch.  This localization
effect is essential in order to obtain a peak in the Wigner function.
The form of the peak and the precise location
of its center are determined by the form of $I_{n,n}(r_2,r_1)$.
To illustrate this better let us assume that the state of the
environment is such that the influence functional can be written in
the form
\be
I_{(n,n)}(r) \simeq e^{i\beta_m(r) \xi^m - {\sigma}^2{\xi_m}{\xi^m}}
\te
where $\beta$ and $\sigma$ are real and $r = {r_1 + r_2\over 2}$,
and we notice that $\sigma$ is related to the degree of
diagonalization of the density matrix.
The Wigner function for this is computed to be
\be
W_{(n)}(r,P) \simeq {C_{(n)}}^2(r)\sqrt{\pi\over {\sigma}^2}
e^{-{(P_m - M{\partial S\over \partial r^m} -\ha\beta_m)}^2}
\te
It is a Gaussian peaked
about the classical trajectory $(P_m - M{\partial S\over \partial
r^m}= 0)$
shifted by a backreaction term $\ha\beta_m$, but with a spread
characterized by $\sigma$, both of these arising from coarse graining
the environment. We also notice a competition between sharp
correlations (related to the sharpness of the above peak) and  the
diagonalization of the density matrix, and this can be formalized
(see Paz and Sinha 1991, 1992) in a set of criteria for the emergence of
classical behavior through a compromise between decoherence and
correlations. It can also be shown  that when the influence
functional is of the form (3.8) the correlations predicted are exactly
those given by the semiclassical Einstein equations. Thus the above
arguments can be tied to the emergence of the semiclassical limit
of quantum fields in curved space time starting from quantum
cosmology. This has been illustrated in Paz and Sinha (1991, 1992) in
the context of various specific cosmological models.

\section{Statistical Mechanics and Quantum Cosmology}

The above two examples give a good illustration of how
models of quantum open systems can be used
to understand some basic issues of quantum cosmology.
The advantage of using statistical mechanical concepts for the
study of issues in quantum cosmology and semiclassical gravity has been
discussed in general terms by Hu (1991)
%
%
who emphasized the importance of the interconnection
between processes of decoherence, correlation, dissipation, noise and
fluctuation, particle creation and backreaction.
(see also Hu, Paz and Zhang 1992, 1993, and Gell-Mann and Hartle 1993).
In our view, these processes are actually different
manifestations of the effect of the environment on the different attributes
of the system (the phase information, energy distribution, entropy content,
etc). Decoherence is related to particle creation as they both can
be related to the Bogolubov coefficients, as is apparent
in the example of Calzetta and Mazzitelli (1991) and Paz and Sinha (1992).
It is also important to consider the nature of noise and fluctuation for
any given environment. In this way their overall effect on the system can be
captured more succintly and effectively by general categorical relations like
the fluctuation-dissipation relations (see Hu 1989, Hu and Sinha 1993).
This also sheds light on how to address questions like defining
gravitational entropy and related questions mentioned in the Introduction
(see Hu 1993b).

So far one has only succeeded in finding a pathway to show how semiclassical
gravity can be deduced from quantum cosmology. One important approximation
which makes this transition possible is the assumption of a WKB wave function.
To see the true colors of quantum gravity, which is nonlinear and is likely
to be also nonlocal, one needs to avoid such simplifications.
One should incorporate dynamical fluctuations both
in the fields and in the geometry without any background field separation,
and deal with nonadiabatic and nonlinear conditions directly.
This is a difficult but necessary task. Calzetta (1991) has tackled the
anisotropy
dissipation problem in quantum cosmology without such an approximation.
Recently Calzetta and Hu (1993) have proposed an alternative approach to
address the quantum to classical transition issue in terms of correlations
between histories. It uses the BBGKY hierarchy truncation scheme to provide
a more natural coarse-graining measure which brings about the decoherence of
correlation histories. This scheme goes beyond a simple system-environment
separation and enables one to deal with nonadiabaticity and nonlinearity
directly as in quantum kinetic theory.

\newpage
\noindent {\bf Acknowledgement}

This essay highlights the work done by Calzetta and the three of us in the
years 1987-1992, a good twenty years after Charlie Misner's seminal works
in relativistic cosmology. It is clear from this coarse sampling how much our
work is indebted to Misner intellectually. It is also our good fortune to
have had personal interactions with Charlie as colleague and friend. We wish
him all the best on his sixtieth birthday and look forward to his continuing
inspiration for a few more generations of relativists.
This work is supported in part by the
National  Science Foundation under grant PHY91-19726.

\noindent {\bf References}
\vskip .5cm

\noindent A. Anderson, Phys. Rev. {\bf D42}, 585 (1990)

\noindent R. Arnowitt, S. Deser, and C. W. Misner, ``The Dynamics of General
Relativity.'' in {Gravitation: An Introduction to Current Research}, ed.
L. Witten, pp. 227-65 (Wiley, New York, 1962)




\noindent V. A. Belinskii, E. M. Lifschitz and I. M. Khalatnikov,
Adv. Phys. {\bf 19}, 525 (1970)

\noindent B. K. Berger, Phys. Rev. {\bf D11}, 2770 (1975)

\noindent B. K. Berger, Ann. Phys. (N. Y.) {\bf 83}, 203 (1974)

\noindent L. Bianchi, {\em Mem. di. Mat. Soc. Ital. Sci.}, {\bf 11}, 267 (1897)

\noindent N. D. Birrell and P. C. W. Davies, {\it Quantum Fields in Curved
Space
   ,}
(Cambridge University Press, Cambridge, 1982).

\noindent A. O. Caldeira and A. J. Leggett, Physica {\bf 121A}, 587 (1983)


\noindent E. Calzetta, Class. Quant. Grav. {\bf 6}, L227 (1989)

\noindent E. Calzetta, Phys. Rev. {\bf D43}, 2498 (1991)

\noindent E. Calzetta and B. L. Hu, Phys. Rev. {\bf D35}, 495 (1987)

\noindent E. Calzetta and B. L. Hu, Phys. Rev. {\bf D37}, 2838 (1988)

\noindent E. Calzetta and B. L. Hu, Phys. Rev. {\bf D40}, 380 (1989)

\noindent E. Calzetta and B. L. Hu, Phys. Rev. {\bf D40}, 656 (1989)

\noindent E. Calzetta and B. L. Hu, "Decoherence of Correlation Histories"
in {\it Directions in General Relativity} Vol 2  (Brill Festschrift),
eds. B. L. Hu, M. P. Ryan and C. V. Vishveshwara (Cambridge Univ. Cambridge,
1993)

\noindent E. Calzetta and F. Mazzitelli, Phys. Rev. {\bf D42}, 4066 (1991)

   )

\noindent P. Candelas and D. W. Sciama, Phys. Rev. Lett. {\bf 38}, 1372 (1977)

\noindent D. M. Capper and M. J. Duff, Nuovo Cimento {\bf A23}, 173 (1975)

\noindent S. Deser, M. J. Duff and C. J. Isham, Nucl. Phys. {\bf B111}, 45
(1976)

\noindent P. C. W. Davies, Rep. Prog. Phys. {\bf 41}, 1313 (1978)

\noindent P. C. W. Davies, L. H. Ford and D. N. Page Phys. Rev. {\bf D34}, 1700
   (1986)

\noindent B. S. DeWitt, Phys. Rev. {\bf160}, 1113 (1967)

\noindent B. S. DeWitt, Phys. Rep. {\bf 19C}, 297 (1975)

\noindent B. S. DeWitt, in {\it Quantum Concepts in Space and Time},
ed. R. Penrose and C. J. Isham (Claredon Press, Oxford, 1986)

\noindent H. F. Dowker and J. J. Halliwell, Phys. Rev. {\bf D46}, 1580 (1992)

\noindent D. Eardley, E. Liang and R. K. Sachs, J. Math. Phys. {\bf 13},
99 (1971)

\noindent G. F. R. Ellis and M. A. H. MacCallum, { Commun. Math. Phys.}
{\bf 12}, 108  (1969)


\noindent R. P. Feynman and F. L. Vernon, Ann. Phys. {\bf 24}, 118 (1963)

\noindent M. V. Fischetti, J. B. Hartle and B. L. Hu,  Phys. Rev.
{\bf D20}, 1757 (1979)

\noindent S. A. Fulling, Phys. Rev. {\bf D7}, 2850 (1973)

\noindent M. Gell-Mann and J. B. Hartle, in {\it Complexity, Entropy and
 the Physics of Information}, ed. W. Zurek, Vol. IX
 (Addison-Wesley, Reading, 1990)

\noindent M. Gell-Mann and J. B. Hartle, Phys. Rev. {\bf D47}  (1993)


   (1988)

\noindent R. Griffiths, J. Stat. Phys. {\bf 36}, 219 (1984)

\noindent A. H. Guth, Phys. Rev. {\bf D23}, 347 (1991)

\noindent S. Habib , Phys. Rev. {\bf D42}, 2566 (1990)

\noindent S. Habib and R. Laflamme, Phys. Rev. {\bf D42}, 4056 (1990)

\noindent J. J. Halliwell, Phys. Rev. {\bf D36}, 3627 (1987)

\noindent J. J. Halliwell, Phys. Rev. {\bf D39}, 2912 (1989)

   ys.{\bf A} (1990)

\noindent J. J. Halliwell, Lectures at the 1989 Jerusalem Winter School, in
{\it Quantum Mechanics and Baby Universes}, eds. S. Coleman, J. Hartle,
T. Piran and S. Weinberg (World Scietific, Singapore, 1991)

\noindent J. J. Halliwell and S. W. Hawking, Phys. Rev. {\bf D31}, 1777 (1985)

\noindent J. B. Hartle, Phys. Rev. Lett. {\bf 39}, 1373 (1977)

\noindent J. B. Hartle, in {\sl Gravitation in Astrophysics},
NATO Advanced Summer Institute, Cargese, 1986 ed. B.  Carter and J. Hartle
(NATO ASI Series B: Physics Vol. 156, Plenum, N.Y 1987).

\noindent J. B. Hartle and S. W. Hawking, Phys. Rev. {\bf D28}, 1960 (1983)

\noindent J. B. Hartle and B. L. Hu,  Phys. Rev. {\bf D20},
1772 (1979)

\noindent J. B. Hartle and B. L. Hu,  Phys. Rev.  {\bf D21}, 2756
(1980)

\noindent
\noindent S. W. Hawking,  {\em Nature} (London),{\bf 248}, 30 (1974)



\noindent O. Heckman and E. L. Schucking, In {\em Gravitation: An Introduction
to Current Research}, ed. L. Witten (Wiley, New York, 1962)

\noindent B. L. Hu, Phys. Lett. {\bf A90}, 375 (1982)

\noindent B. L. Hu, Phys. Lett. {\bf A97}, 368 (1983)

\noindent B. L. Hu, in {\it Cosmology of the Early Universe}, ed. L. Z. Fang
and R. Ruffini (World Scientific, Singapore, 1984)

\noindent B. L. Hu, Physica {\bf A158}, 399 (1989)

\noindent B. L. Hu "Quantum and Statistical Effects in Superspace Cosmology"
in {\it Quantum Mechanics in Curved Spacetime}, ed. J. Audretsch
and V. de Sabbata (Plenum, London 1990)

\noindent B. L. Hu, in {\it Relativity and Gravitation: Classical
and Quantum}, Proc. SILARG VII, Cocoyoc, Mexico 1990,
eds. J. C. D' Olivo et al (World Scientific, Singapore 1991)

\noindent B. L. Hu, "Statistical Mechanics and Quantum Cosmology",
in {\it Proc. Second International Workshop on Thermal Fields and Their
Applications}, eds. H. Ezawa et al (North-Holland, Amsterdam, 1991)

\noindent B. L. Hu, "Fluctuation, Dissipation and Irreversibility"
in {\it The Physical Origin of Time-Asymmetry}, Huelva, Spain, 1991,
eds. J. J. Halliwell, J. Perez-Mercader and W. H. Zurek
(Cambridge University Press, 1993)



\noindent B. L. Hu, "Quantum Statistical Processes in the Early Universe"
in {\it Quantum Physics and the Universe}, Proc. Waseda Conference, Aug. 1992
ed. Namiki, K. Maeda, et al  (Pergamon Press, Tokyo, 1993)

\noindent B. L. Hu and H. E. Kandrup, Phys. Rev. {\bf D35}, 1776 (1987)

\noindent B. L. Hu and L. Parker, Phys. Lett.  {\bf 63A}, 217 (1977)

\noindent B. L. Hu and L. Parker, Phys. Rev. {\bf D17}, 933 (1978)

\noindent B. L. Hu and D. Pavon, Phys. Lett. {\bf B180}, 329 (1986)


\noindent B. L. Hu, J. P. Paz and Y. Zhang, Phys. Rev. {\bf D45}, 2843 (1992)

\noindent B. L. Hu, J. P. Paz and Y. Zhang,
Phys. Rev. {\bf D47}  (1993)

\noindent B. L. Hu, J. P. Paz and Y. Zhang, ``Stochastic Dynamics of
Interacting Quantum Fields'' Phys. Rev. D (1993)


\noindent B. L. Hu and S. Sinha, "Fluctuation-Dissipation Relation in
Cosmology"
     Univ. Maryland preprint (1993)



\noindent B. L. Hu and Y. Zhang, "Coarse-Graining, Scaling, and Inflation"
Univ. Maryland Preprint 90-186 (1990)

\noindent E. Joos and H. D. Zeh, Z. Phys. {\bf B59}, 223 (1985)

\noindent R. D. Jordan, Phys. Rev. {\bf D33}, 44 (1986)



\noindent H. E. Kandrup, Phys. Rev. {\bf D37}, 3505 (1988)

\noindent L. V. Keldysh, Zh. Eksp. Teor. Fiz. {\bf 47}, 1515 (1964)
[Sov. Phys. JEPT {\bf 20}, 1018 (1965)]

\noindent C. Kiefer, Class. Quant. Grav. {\bf 4}, 1369 (1987)

\noindent C. Kiefer, Class. Quant. Grav. {\bf 8}, 379 (1991)

\noindent K. Kuchar, Phys. Rev. {\bf D4}, 955 (1971)

\noindent K. Kuchar and M. P. Ryan Jr., in {\em Proc. of Yamada Conference
XIV} ed. H. Sato and T. Nakamura (World Scientific, 1986)

\noindent K. Kuchar and M. P. Ryan Jr.,  Phys. Rev.  {\bf D40}, 3982 (1989)

\noindent R. Laflamme and J. Luoko, Phys. Rev. {\bf D43}, 3317(1991)

   and W. Zwerger, Rev. Mod. Phys. 59 No. 1 (1987) 1


\noindent E. M. Lifschitz and I. M. Khalatnikov, Adv. Phys. {\bf 12}, 185
(1963)

\noindent V. N. Lukash and A. A. Starobinsky, JETP {\bf 39}, 742 (1974)

\noindent R. A. Matzner and C. W. Misner, Ap. J. {\bf 171}, 415 (1972)

\noindent C. W. Misner, Ap. J. {\bf 151}, 431 (1968)

\noindent C. W. Misner, Phys. Rev. Lett. {\bf 22}, 1071 (1969)

\noindent C. W. Misner, Phys. Rev. {\bf186}, 1319 (1969)

\noindent C. W. Misner, in {\em Magic Without Magic}, ed. J. Klauder
(Freeman, San Francisco, 1972)

\noindent E. Mottola, Phys. Rev. {\bf D33}, 2136 (1986)

\noindent R. Omnes, J. Stat. Phys. {\bf 53}, 893, 933, 957 (1988);
Ann. Phys. (NY) {\bf 201}, 354 (1990): Rev. Mod. Phys. {\bf 64}, 339 (1992)

\noindent T. Padmanabhan, Phys. Rev. {\bf D39}, 2924 (1989)


\noindent T. Padmanabhan and T. P. Singh, Class. Quan. Grav. {\bf 7}, 411
(1990)

\noindent L. Parker, ``The Creation of Particles in an Expanding Universe,''
Ph.D. Thesis, Harvard University (unpublished, 1966).

\noindent L. Parker, { Phys. Rev. } {\bf 183}, 1057 (1969)

\noindent L. Parker, { Phys. Rev. } {\bf D7}, 976 (1973)

\noindent J. P. Paz, Phys. Rev. {\bf D41}, 1054 (1990)

\noindent J. P. Paz, Phys. Rev. {\bf D42}, 529 (1990)

   .

\noindent J. P. Paz and S. Sinha, Phys. Rev. {\bf D44}, 1038 (1991)

\noindent J. P. Paz and S. Sinha, Phys. Rev. {\bf D45}, 2823 (1992)

\noindent J. P. Paz and W. H. Zurek, in preparation (1993)

\noindent R. Penrose, ``Singularities and Time-Asymmetry'' in {\em
General Relativity: an Einstein Centenary Survey}, eds. S.W. Hawking and
W. Israel (Cambridge University Press, Cambridge, 1979)


\noindent M. P. Ryan, {\it Hamiltonian Cosmology} (Springer, Berlin, 1972)

\noindent M. P. Ryan, Jr.  and L. C. Shepley, {\em Homogeneous Relativistic
Cosmologies} (Princeton University Press, Princeton, 1975)

\noindent J. Schwinger, { Phys. Rev.}, {\bf 82}, 664 (1951)

\noindent J. S. Schwinger, J. Math. Phys. {\bf 2}  407 (1961).


\noindent R. U. Sexl and H. K. Urbantke, { Phys. Rev.}, {\bf 179}, 1247
(1969)


\noindent S. Sinha, Ph. D. Thesis, University of Maryland (unpublished 1991)

\noindent S. Sinha and B. L. Hu, Phys. Rev. {\bf D44}, 1028 (1991).



\noindent A. H. Taub, { Ann. Math.}, {\bf 53}, 472 (1951)


\noindent T. Vachaspati and A. Vilenkin, Phys. Rev. {\bf D37}, 898 (1988)

\noindent A. Vilenkin, Phys. Rev. {\bf D27}, 2848 (1983), {\bf D30}, 509
(1984);
Phys. Lett. {\bf 117B}, 25 (1985)

\noindent J. A. Wheeler, in {\it Relativity, Groups and Topology}, ed.
B. S. De Witt and C. De Witt (Gordon and Breach, New York, 1964)

\noindent  J. A. Wheeler, in {\it Battelle Rencontres}
ed. C. DeWitt and J.A Wheeler (Benjamin, N. Y. 1968)

\noindent J. A. Wheeler and W. H. Zurek, {\it Quantum Theory and Measurement}
(Princeton University Press, Princeton 1986)

\noindent E. P. Wigner, Phys. Rev. {\bf 40}, 749 (1932)


\noindent H. D. Zeh, Phys. Lett. {\bf A116}, 9 (1986).

\noindent Ya. B. Zel'dovich, Pis'ma Zh. Eksp. Teor. Fiz, {\bf 12} ,443 (1970)
[JETP Lett. {\bf 12}, 307(1970)]

\noindent Ya. B. Zel'dovich, in {\it Physics of the Expanding Universe},
ed. M. Diemianski (Springer, Berlin, 1979)

\noindent Ya. B. Zel'dovich and A. A. Starobinsky,
Zh. Teor. Eksp. Fiz. {\bf 61} (1971)  2161 [JEPT {\bf 34}, 1159 (1972)]


\noindent W. H. Zurek, Phys. Rev. {\bf D24}, 1516 (1981); {\bf D26}, 1862
(1982); in {\it Frontiers of Nonequilibrium Statistical Physics},
ed. G. T. Moore and M. O. Scully (Plenum, N. Y., 1986);
Physics Today {\bf 44}, 36 (1991)



\noindent R. Zwanzig, in {\it Lectures in Theoretical Physics},
ed. W. E. Britten, B. W. Downes and J. Downes (Interscience, New York, 1961)

\end{document}